\documentstyle[11pt,newpasp,twoside,epsfig,epsf]{article}
\def\edcomment#1{\iffalse\marginpar{\raggedright\sl#1\/}\else\relax\fi}
\reversemarginpar

\begin{document}
\title{ FLIERs as stagnation knots from partially collimated outflows}
 \author{Wolfgang Steffen }
\affil{Instituto de Astronom\'{\i}a y Meteorolog\'{\i}a, Universidad de
Guadalajara,  Av. Vallarta 2602, 44130 Guadalajara, Jal., M\'exico }
\author{Jos\'e Alberto L\'opez }
\affil{Instituto de Astronom\'{\i}a, Universidad Nacional Aut\'onoma de
M\'exico,  Apartado Postal 877, 22800 Ensenada, B.C., M\'exico}

\begin{abstract}
We propose a new model for the formation of fast, 
low-ionization emission regions (FLIERs) in planetary nebulae
that is able to account for many of their attendant characteristics
and circumvent the problems on the collimation/formation mechanisms found
in previous studies. In this model, FLIERs are formed in the stagnation 
zone of partially collimated stellar winds with reduced momentum flow 
along the axis. A concave bow-shock structure is formed due to the lack of
momentum flow along the axis of a midly bipolar stellar wind.
The stagnation knots are formed when the shocked environment
medium accumulates at the apex of the outer shell and is compressed 
to a dense knot in the {\em concave} section of the bow-shock. 
We present  two-dimensional hydrodynamic simulations of the formation of 
a stagnation knot and compare the resultant dynamical properties with 
those of FLIERs in planetary nebulae. 

\end{abstract}

\section{Introduction and model description}

FLIERs in planetary nebulae were originally identified with the structures
previously known as ansae in elliptical planetary nebulae (Balick et al. 1993)
though their peculiar characteristics are now recognized in a much wider
variety of PNe (e.g. Corradi et al. 1996. Guerrero et al. 1999), and their 
interpretation has resisted a consistent explanation up to date (Balick et al.
1998). FLIERs are characterised by outflow radial velocities of the order of
30-50 km/s; ionzation gradients decreasing outwards from the nebular core
and  `head-tail' morphologies, notably with the tails pointing outwards from
the nucleus. More recent observations reveal very high radial velocities for
other FLIERs (Corradi et al. 1999, Gon{\c c}alves et al.
these proceedings, Redman et al. 1999). Mainly two types of models have been
explored for the formation  of FLIERs (see Balick et al. 1998 and references
therein), namely, ionization  fronts (IF) on localized dense knots or
bow-shocks of fast knots ramming  through the shell of the PN. Recently,
Redman \& Dyson (1999) and Dyson \& Redman (these proceedings) discuss a model
in which FLIERs represent recombination fronts (RF) in mass-loaded jets. 

In this work we present a simple hydrodynamic model for the formation of 
symmetric axial knots with  supersonic velocities from a 
low-density bipolar outflow with reduced momentum flux along its axis. 
The knots are formed from cooling shocked ambient gas in the stagnation 
region of the combined outflow. The evolution of a dense knot propagating 
through a thin ambient medium has been studied by Jones, Kang \& Tregillis
(1994) using  hydrodynamic simulations and by Soker \& Regev (1998) from
analytic arguments in the specific context of FLIERs. However, 
those works do not discuss the formation process for the knots.
Previously, we introduced the idea of a "stagnation knot" to model the large
scale structure of the giant envelope of the PN KjPn8 (Steffen \& L\'opez,
1998). The reduced momentum flux around the axis causes the bow-shock to become
concave instead of convex in this region. If the bow-shock is non-radiative,
the ambient medium passing through the oblique region of the shock is then
"refracted" towards the axis, instead of away from it, as it happens in
conventional bow-shocks. The accumulated material in the stagnation region may
then be held together for sufficient time to cool and be compressed to a dense
knot. If the shock is radiative, the compressed post-shock material is later
crushed to a single or multiple knots on and around the axis. As long as the
central outflow continues and drives the expanding shock, the stagnation knot
will move roughly at the same speed as the rest of the bow-shock and remains at
the bright rim formed by shocked ambient gas. However, when the outflow ceases
the envelope slows down rather quickly whereas the dense knot continues to move
outwards at its original speed. As the amount of swept up mass from the ambient
medium increases, the knot's motion is progressively slowed down.
\section{Simulations}
            
In order to investigate the dynamical properties of the knots formed in the
stagnation region of a partially collimated low-density outflow, we present 
two cases (A \& B) of 2D-hydrodynamical simulations in axisymmetry using the
Corali-code (Raga et al. 1995) with a 5-level binary adaptive grid and a 
maximum of $513\times257$ grid cells with a physical sizes of $5\times10^{17}$~cm 
by $2.5\times10^{17}$~cm for run A and half of these
measures for run B. The cooling  was calculated according to the description
in Steffen  et al. (1997) and references therein. The outflow was initialized
on a sphere with a radius of $2.5\times10^{16}$~cm ($2.5\times10^{16}$~cm), a
velocity of 2000~km/s (800~km/s) and a half opening angle $\theta_0=15$\deg
(15\deg) (the angle of highest momentum flux) for run A (B). The initial number
density of the outflow is $30~{\rm cm}^{-3}$ ($40~{\rm cm}^{-3}$) constant on 
the sphere, whereas that of the ambient medium was assumed to be $120~{\rm
cm}^{-3}$. The outflow is switched-off after $1.5\times10^{10}$ seconds (473
years) for both runs. The momentum flux was modulated as a  function of the
angle $\theta$ from the axis arbitrarily using
$v(\theta)=v_0(\theta_0/\theta)^2$ for $\theta>\theta_0$ and
$v(\theta)=v_0/(1-0.25(\theta-\theta_0)/\theta)$ for $\theta<=\theta_0$. S

\section{Results}

Key ingredients often found in PNe with FLIERs are observed
to form in the models. The outflow creates a low density bipolar cavity 
with an outer dense and bright rim (see Figure 1) of shocked halo
gas which propagates at a few tens of kilometers per second. The rim develops
instabilities which produce high density knots propagating at velocities similar
to the rim. These knots might be identified as non-axial FLIERs similar to those
observed in NGC 7662 (see also Dwarkadas \& Balick 1998). In run A 
(Figure 1a to 1d) after forming a sort of dense "polar cap" at $t\approx 300$~years 
the stagnation region begins to be compressed to a knot, which we
associate with the FLIER. The expansion speed of the rim, the stagnation knot
and the instability knots ranges between 60 and 150~km/s at this time. After
the stellar wind ceases the expansion speed of the rim and the instability
knots rapidly drops to around 50~km/s.  The stagnation knot, however,
continues its linear motion at a speed of around 150~km/s which drops more
slowly. This model leads to representative parameters consistent with
observations of PNe with FLIERs.

In the more extreme case of run B (Figure 1e and 1f) the stagnation knot
reaches 250~km/s and produces a long feature of fast material far away 
from the main nebula. The region through which the stagnation knot has
propagated shows an outward increase in velocity (Fig. 1d,f). As the dense 
knot continues, the smaller pieces spread out along the path developing
a kinematic pattern of roughly linear increase of speed with distance, as
observed in MyCn 18 (Bryce et al. 1997, Redmann et al., these proceedings) and
other recent cases (Corradi et al. 1999, and Gon{\c c}alves these proceedings).

Within the framework of this model, it would be interesting to search for 
those young elliptical PNe that show signs of polar caps as likely candidates
to develop into compact FLIERs. Full details of this work will appear shortly
elsewhere.

\begin{figure}
\plotone{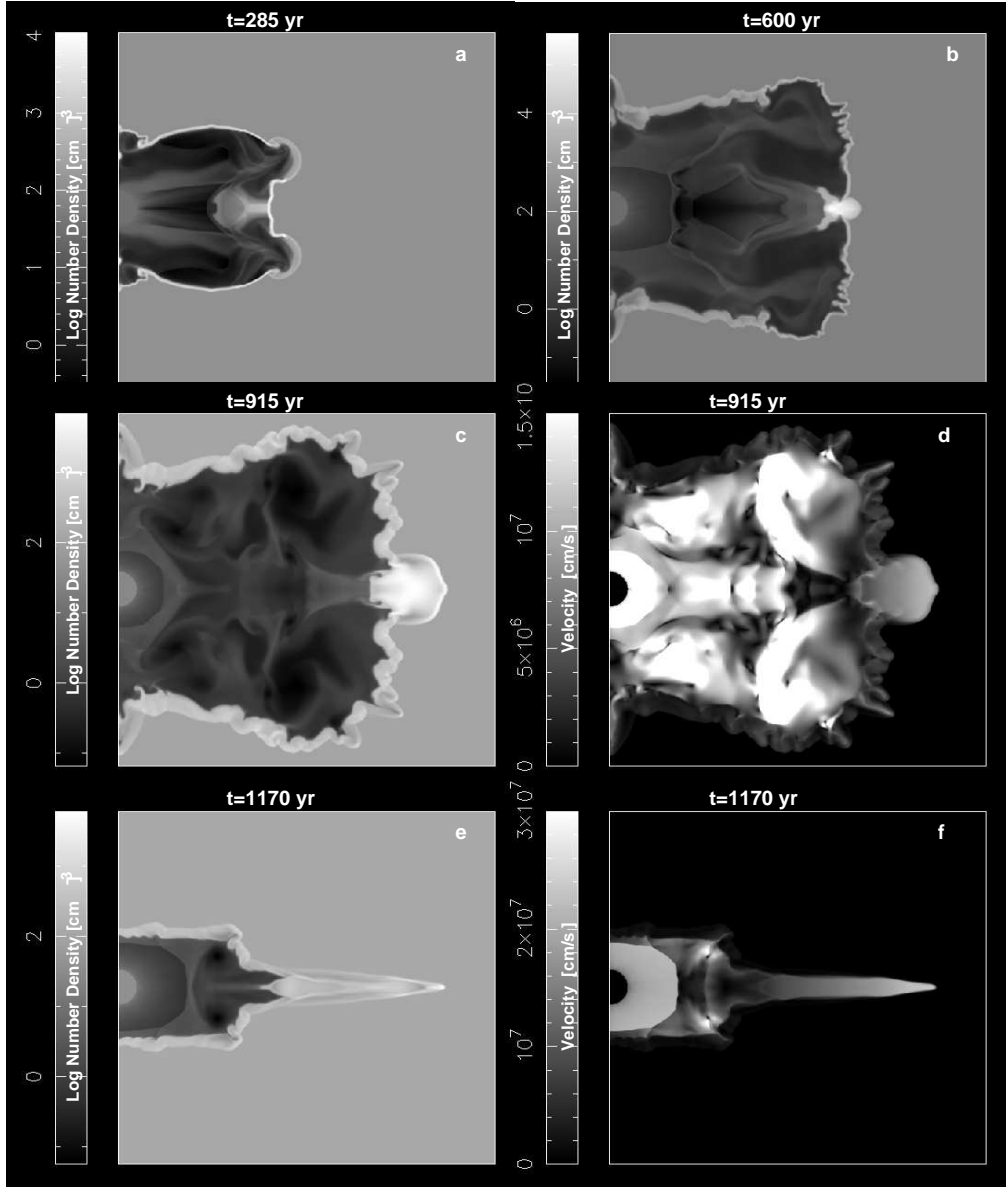}
\caption{A sequence of density cuts through run A (a-c) and run B (e)
as described in the text. Panels "d" and "f" represent the velocity fields
corresponding to the densities in "c" and "e".}
\end{figure}

\end{document}